\documentclass[11pt,english]{article}
\usepackage{mathptmx}
\usepackage[T1]{fontenc}
\usepackage[latin9]{inputenc}
\usepackage[letterpaper]{geometry}
\usepackage{color}
\usepackage{float}
\usepackage{amstext}
\usepackage{graphicx}
\usepackage{setspace}
\makeatletter

\newcommand{\lyxline}[1][1pt]{%
  \par\noindent%
  \rule[.5ex]{\linewidth}{#1}\par}
\providecommand{\tabularnewline}{\\}

\newcommand{\lyxaddress}[1]{
\par {\raggedright #1
\vspace{1.4em}
\noindent\par}
}

\author{}
\date{}
\usepackage{cite}

\@ifundefined{showcaptionsetup}{}{%
 \PassOptionsToPackage{caption=false}{subfig}}
\usepackage{subfig}
\makeatother

\usepackage{babel}

\begin{document}
\begin{singlespace}

\title{{\Large Self-organized synthesis of patterned magnetic nanostructures
with in-plane and perpendicular to the plane magnetization }}
\end{singlespace}

\begin{singlespace}

\date{H. Krishna$^{\text{1}}$, %
\thanks{Contact author: anup@wuphys.wustl.edu%
} A. K. Gangopadhyay$^{\text{1}}$, J. Strader$^{\text{2}}$, and %
\thanks{Contact author:ramki@utk.edu%
} R. Kalyanaraman$^{\text{2,3}}$}
\end{singlespace}

\maketitle
\begin{singlespace}

\lyxaddress{\begin{center}
{\small $^{\text{1}}$Department of Physics and Center for Materials
Innovation, Washington University in St. Louis, MO 63130}
\par\end{center}}

\lyxaddress{\begin{center}
{\small $^{\text{2}}$Department of Materials Science and Engineering,
University of Tennessee, Knoxville, TN 37996}
\par\end{center}}

\lyxaddress{\begin{center}
{\small $^{\text{3}}$Department of Chemical and Biomolecular Engineering,
University of Tennessee, Knoxville, TN 37996}
\par\end{center}}
\end{singlespace}
\begin{abstract}
Patterned arrays of ferromagnetic nanoparticles of Co, Ni, and Fe$_{\text{50}}$Co$_{\text{50}}$
have been synthesized from their ultrathin metal films on SiO$_{\text{2}}$
substrate by nanosecond laser-induced self-organization. The morphology,
nanostructure, and magnetic behavior of the nanoparticle arrays were
investigated by a combination of electron, atomic force, and magnetic
force microscopy techniques. Transmission electron microscopy investigations
revealed a granular polycrystalline nanostructure, with the number
of grains inside the nanoparticle increasing with their diameter.
Magnetic force measurements showed that the magnetization direction
of the Co and Ni nanoparticles was predominantly out-of-plane while
those for the Fe$_{\text{50}}$Co$_{\text{50}}$ alloy was in the
plane of the substrate. This difference in behavior is due to the
dominating influence of magnetostrictive energy on the magnetization
as a result of residual thermal strain following fast laser processing.
Since the magnetostriction coefficient is negative for polycrystalline
Co and Ni, and positive for Fe$_{\text{50}}$Co$_{\text{50}}$, the
tensile residual strain forces the magnetization direction of the
negative magnetostriction materials out-of-plane and the positive
magnetostriction materials in-plane. This demonstrates a cost-effective
non-epitaxial technique for the fabrication of patterned arrays of
magnetic nanoparticles with tailored magnetization orientations.

\lyxline{\small}
\end{abstract}

\section*{1. Introduction\label{sec:1.-Introduction}}

Fabrication of patterned nanostructures consisting of discrete nanoparticles
whose physical properties (\textit{e.g.} magnetic, semiconducting,
optical) can be reliably controlled by shape, size, and spacing, in
conjunction with processing parameters, is of prime importance in
the field of nanotechnology. Arrays of magnetic nanoparticles can
be used for many applications, including high density magnetic data
storage \cite{Todorovic99} to non-volatile and high speed magnetic
random access memories (MRAM) \cite{Chou96}, opto-electronics \cite{Saleno02},
and biological sensor applications \cite{Dave09,Molday82}. One of
the important challenges in the applications of magnetic nanoparticles
is the control of magnetic orientation of each nanoparticle, which
can provide additional advantages. For example, particles with perpendicular
to the plane anisotropy enable higher density for the same signal-to-noise
ratio \cite{Bertram00} and lower read and write errors \cite{Albrecht03},
compared to particles with in-plane magnetization. In the absence
of an external magnetic field, the magnetic moment of a ferromagnetic
material aligns spontaneously along a preferred direction. This direction
corresponds to the minimum magnetic energy, which is determined by
intrinsic material parameters such as crystalline anisotropy, as well
as extrinsic parameters related to the processing conditions, such
as shape, size, and strain. Therefore, achieving desired magnetic
orientation and switching behavior requires control of magnetic energy
through the choice of materials and processing parameters. For instance,
epitaxial thin film growth technique can be used to synthesize single
crystal magnetic nanodots, where the magnetocrystalline anisotropy
uniquely determines the magnetization direction \cite{Hwang06}. 

In this work, we demonstrate magnetic orientation control in polycrystalline
nanoparticles through non-epitaxial means. We have synthesized ordered
magnetic nanoparticle arrays with the magnetization direction tailored
either in-plane or perpendicular to the plane. The magnetic nanoparticle
arrays were produced on amorphous SiO$_{2}$ surfaces by nanosecond
(ns) laser-induced self organization of nm thick ferromagnetic metal
films. In this process, the ns laser pulse melts the film, which then
undergoes a spontaneous change in morphology. Application of multiple
pulses leads to self-organized nanoparticles with predictable particle
size and interparticle spacing. The self-organization is a result
of spinodal dewetting \cite{KrishnaPCCP09,FavazzaAPL06,FavazzaNano06}
or thermocapillary driven flow \cite{Favazza07,TricePRL08,Gangopadhyay07}.
We have used this approach previously to synthesize single-domain
Co \cite{krishna08a} and Fe nanoparticles \cite{KrishnaSPIE08},
which showed particle size-dependent magnetic anisotropy behavior.
Here we have applied this technique to orient the magnetization direction
of the nanoparticles either in-plane or perpendicular to the substrate
plane. We have chosen two elemental ferromagnets with negative magnetostriction
coefficient, $\lambda_{S}$, ($\lambda_{S}$= -30 ppm for Co and $\lambda_{S}$
= -34 ppm for Ni) and one alloy, Fe$_{\text{50}}$Co$_{\text{50}}$,
with positive magnetostriction coefficient ($\lambda_{S}$ = +84 ppm)
\cite{CullityBook,Hall59,Lorenz04}; the quoted values of $\lambda_{S}$
are for polycrystalline materials. The ensuing investigations of magnetic
properties show that single-domain nanoparticles of Co and Ni have
preferential perpendicular (out-of-plane) magnetic orientation, while
those of FeCo have preferential in-plane orientation. The reason for
this difference was attributed to the coupling of strain (tensile),
generated within the nanoparticles by the substrate during rapid thermal
processing, to the magnetostriction. The opposite sign of the magnetostriction
coefficients for Co and Ni compared to FeCo is responsible for their
different orientations.

\section*{2. Experimental Details\label{sec:2.-Experimental-Details}}

A thin film of Co ($\sim4$ nm) was deposited using electron beam
evaporation (e-beam), while Ni ($\sim5$ nm), and Fe$_{\text{50}}$Co$_{\text{50}}$
($\sim4$ nm) films were deposited using pulsed laser deposition (PLD)
technique on commercially available optically smooth SiO$_{\text{2}}$/Si(100)
substrates under ultra high vacuum ($\sim1\times10^{-8}$ Torr). The
thermally grown oxide (SiO$_{\text{2}}$) layer was 400 nm thick.
Prior to evaporation, the samples were cleaned in an ultrasonic bath
with acetone, followed by methanol and de-ionized water. The ingot
for PLD used for FeCo alloy was made by repeated arc-melting of a
stoichiometric mixture of Co and Fe (4N pure, Alfa Aesar). The arc-melting
was performed in a water-cooled copper hearth under a high purity
TiZr-gettered argon atmosphere. The thicknesses of the films were
determined by in-situ quartz crystal thickness monitor and calibrated
electron dispersive x-ray spectroscopy (EDS) measurements. The films
were irradiated with a Nd:YAG pulsed laser beam (266 nm wavelength,
9 ns pulse width, and 50 Hz repetition rate) under high vacuum. The
energy density of the laser pulses ($\sim100$ mJ/cm$^{\text{2}}$)
was chosen to be slightly above the melt threshold \cite{Trice07}.
Approximately 3000 laser pulses were required to achieve the pattern
with arrays of nanoparticles. Two types of laser irradiation experiments
were performed. For the case of Ni and FeCo, a spatially uniform single
beam was incident perpendicular to the substrate surface to produce
nanoparticle arrays by spinodal dewetting \cite{KrishnaPCCP09,FavazzaAPL06,FavazzaNano06}.
For Co, two beam laser-interference irradiation was performed to produce
a 1-dimensional ordered nanoparticle array \cite{Favazza07}. 

The resulting nanoparticle arrays were characterized by scanning electron
microscopy (SEM, Hitachi S-4500) and by transmission electron microscopy
(TEM, JEOL 2100F) using a 200 KeV beam. The TEM samples were prepared
by a chemical etching method \cite{Lydia06}. Tapping mode atomic
force microscopy (AFM) and zero-field magnetic force microscopy (MFM),
using a Digital Instruments Dimension 3000 instrument, were performed
on the nanoparticle arrays to obtain the topographic image, and magnetization
direction, respectively. A silicon cantilever, coated with a few tens
of nm thick CoCr alloy (Asylum Research, ASY), was used in the MFM
measurements at a scan height of 50 nm. To rule out any influence
of the MFM-tip on the measurement, MFM was performed in different
directions ($0^{o}$ and $90^{o}$) and at different heights ($20$,
$50$ and $100$ nm) from the sample. The orientation of the magnetization
of individual nanoparticles was determined by comparing the MFM image
contrast of the nanoparticles with simulated images \cite{krishna08a}.

To determine the residual thermal strain, finite element simulations
were performed on $20$ - $100$ nm diameter hemispherical nanoparticle
on top of $400$ nm thick SiO$_{2}$ substrate under the processing
conditions. This was accomplished using the COMSOL software package
for a 2-D axi-symmetric geometry, where the particle is perfectly
adhesed to the substrate. The model is set to be stress-free at the
melting point of the nanoparticle, and the stresses and strains generated
during cooling to room temperature due to thermal contraction are
determined. The substrate is assumed to be elastic, and the particle
is modeled as both a purely elastic (in which case the yield stress
is infinite) and elastic-perfectly plastic (in which case the standard
yield strength, $\sigma_{y}$, of the material is used)\textcolor{red}{{}
}solid \cite{Dieter}. The yield strength, $\sigma_{y}$, of annealed
pure Co (400 MPa) and pure Ni (300 MPa) was determined by converting
large depth hardness data by means of the tabor relation, Hardness
= 3$\sigma_{y}$ \cite{Karimpoor03,Zong06,TaborBook}. The average
stresses within the particle were converted to elastic strain by Hooke's
law, resulting in two tensile in-plane principal elastic strains,
corresponding to the radial and hoop strains (in cylindrical geometry)
\cite{Dieter}. The various materials parameters used and results
obtained are tabulated in table \ref{tab:Table}.

\section*{3. Results\label{sec:3.-Results}}

The microstructural studies on the arrays of Co, Ni and Fe$_{\text{50}}$Co$_{\text{50}}$
particles are shown in Fig.\ref{fig:Bright-field-TEM}. Figure\ref{fig:Bright-field-TEM}(a)
is the bright field (BF) TEM micrograph of Co nanoparticles. Detailed
microstructural analysis for Co particles, published in ref. \cite{krishna08a},
revealed a granular microstructure with random orientation of the
grains inside the nanoparticles. The number of grains increased with
increasing nanoparticle size from 1 grain (\textit{i.e}. single crystal)
for the smaller particles ($<40$ nm) to 20-30 grains for the bigger
particles ($\sim120$ nm), with small statistical variation when different
similar size particles were compared. Figure\ref{fig:Bright-field-TEM}(b)
is the BF TEM image for the array of Ni nanoparticles. These also
indicate a granular nanostructure with random grain orientation. Similar
to Co, the very small Ni particles ($\sim15$ nm) are single grained
while the bigger particles have multiple grains (\emph{e.g.}, $\sim40$
nm particles show nearly 20 grains). The TEM image of Fe$_{\text{50}}$Co$_{\text{50}}$
nanoparticles {[}shown in Fig.\ref{fig:Bright-field-TEM}(c){]} also
showed similar granular behavior. The main difference was the large
statistical variation in the number of grains from particle to particle,
even when the size was similar (\emph{e.g.}, a small fraction of $\sim150$
nm size particles had only 2-5 grains, while the majority had more
than 15 grains).

Figure\ref{fig:Co-AFM-MFM} shows the AFM (\ref{fig:Co-AFM-MFM}(a))
and zero-field MFM (\ref{fig:Co-AFM-MFM}(b)) images of an array of
Co nanoparticles produced by two beam irradiation. The separation
between the rows of particles is $400$ nm, consistent with the separation
of the interference fringes from the two beams. The regular 1D pattern
of the nanoparticles along the lines is clearly evident. A 2D pattern
can also be formed using three beam irradiation \cite{Gangopadhyay07}.
The average particle diameter was measured to be $110\pm34$ nm. The
corresponding MFM image in Figure \ref{fig:Co-AFM-MFM}(b) shows that
the image contrast of the particles is either uniformly dark with
a bright periphery or uniformly bright with a dark periphery. When
compared with the simulated MFM image contrast of single domain particles
with different magnetization directions (Fig.\ref{fig:MFM-simulation}),
it is clear that all particles in Fig.\ref{fig:Co-AFM-MFM}(b) are
single domain and have their magnetization oriented perpendicular
to the substrate plane; the exactly opposite image contrast of the
two groups is due to their magnetization pointing either up or down.
The single domain behavior of such large diameter particles, exceeding
that of theoretically calculated single domain size of 60 nm for single
crystal Co particles \cite{Kittel46,CullityBook}, is possibly due
to strong exchange coupling among the grains in these polycrystalline
nanoparticles.

The AFM and MFM images for the Ni array produced by a single beam
irradiation, is shown in Figure\ref{fig:Ni-AFM-MFM}(a) and \ref{fig:Ni-AFM-MFM}(b),
respectively. Due to self-organization by spinodal dewetting, the
particles have a characteristic interparticle spacing (\textasciitilde{}
615 nm) and a fairly narrow particle size distribution ($176\pm37$
nm). Compared to the two beam irradiation, the spatial distribution
of these particles do not follow any pattern, however. Similar to
Co, the contrast in the MFM image {[}Fig.\ref{fig:Ni-AFM-MFM}(b){]}
indicates that almost all particles (similar in size to that marked
as \# 3) have their magnetization perpendicular to the substrate plane,
either up or down. A few particles (\textit{\textcolor{black}{e.g.}}
marked as 2 in Figure\ref{fig:Ni-AFM-MFM}(a)) have their magnetization
at an angle $<90^{o}$ to the plane; only the smallest particle (marked
as 1) of about 75 nm in diameter is oriented at a small angle to the
plane. Again, the nanoparticles have a multi-grained microstructure,
but are single domain up to about 220 nm diameter, which is slightly
larger than the previously reported value of \textasciitilde{} 180
nm for spherical single grain Ni nanoparticle \cite{Liu06}. 

The AFM and the MFM images on an array of Fe$_{\text{50}}$Co$_{\text{50}}$
nanoparticles, produced by a single beam irradiation of a $\sim4$
nm film are shown in Figures\ref{fig:FeCo-AFM-MFM}(a) and (b), respectively.
The average particle diameter is $113\pm32$ nm with about \textasciitilde{}
580 nm separation. The particles with diameters $50$ nm and $150$
nm, indicated as $1$ and $2$ in the AFM and MFM images, show in-plane
($0^{o}$) magnetization. In stark contrast to Co and Ni nanoparticles,
most of the Fe$_{\text{50}}$Co$_{\text{50}}$ nanoparticles (around
70 \%) show in-plane magnetization while the rest (\textit{\textcolor{black}{e.g.}}
number 3) are at a small angle ($\leq45^{o}$) to the plane. These
multigrain particles remain single domain up to about $175$ nm in
diameter. These results clearly show a difference in the orientation
of the magnetization of the nanoparticles with respect to the substrate
plane when the magnetostriction coefficient changes sign.

The average in-plane elastic thermal strains determined from the finite
element simulations were found to be independent of particle size
since the only length scale present in the problem is the ratio of
particle diameter to the size of the SiO$_{2}$ substrate, which is
sufficiently large. The larger of the two (radial versus hoop strain)
average in-plane elastic strains is given in Table\ref{tab:Table},
along with the material parameters used in the simulation \cite{Alers60}.
Only the average elastic component of strain is reported due to its
contribution to the magnetostrictive energy. With no mechanism to
relieve internal stresses, the elastic solutions (corresponding to
the columns with $\infty$ yield stress in table \ref{tab:Table})
represent an upper bound on the average elastic strains. The average
elastic strain is found to be lower for the elastic-perfectly plastic
simulations (corresponding to the columns with finite values of yield
stress in table \ref{tab:Table}) due to the presence of large plastic
deformations. The yield strength chosen and the nature of the elastic-perfectly
plastic simulations neglects any size, strain hardening and cooling
rate affects, which may significantly reduce plastic deformations
consequently increasing average stresses and elastic strains within
the particle \cite{Dieter,MaClarke}. For this reason, the elastic-plastic
simulations are assumed to be a lower bound of the resulting average
elastic strain.

\section*{4. Discussion}

To understand this difference, we now focus on the various contributions
to the magnetic energy of a nanoparticle. The magnetocrystalline anisotropy,
which depends on the crystal structure, is significant only for single
crystal or polycrystalline particles with preferred crystallographic
orientation of the grains. For random crystallographic orientation
of the grains, such as the case here, the contribution of crystalline
anisotropy scales inversely with the number of grains \cite{New95}.
We have estimated that the crystalline anisotropy is at least one
to two orders of magnitude smaller than the single crystal value for
nanoparticles that contain more than 20 grains (see Fig.\ref{fig:Energy-Phase-plots}
and \cite{krishna08a}). Because of the large interparticle separation
(400 to 600 nm), dipolar interaction energy is also small ($\leq10$
J/m$^{\text{3}}$). The shapes of these nanoparticles are nearly hemispherical
as has been determined by the AFM measurements. The estimated contact
angles were $104\pm22^{o}$, $106\pm26^{o}$, and $103\pm20^{o}$,
for Co, Ni, and Fe$_{\text{50}}$Co$_{\text{50}}$, respectively.
For the average particle size and separations, using the known saturation
magnetizations (1400, 485, and 1922 Gauss for Co, Ni, and Fe$_{\text{50}}$Co$_{\text{50}}$,
respectively \cite{CullityBook}), the demagnetization energy was
estimated to be $\sim1.8\times10^{3}$, $\sim2.3\times10^{2}$, and
$\sim3.6\times10^{3}$ J/m$^{\text{3}}$ for Co, Ni, and Fe$_{\text{50}}$Co$_{\text{50}}$,
respectively (see horizontal lines in Fig.\ref{fig:Energy-Phase-plots}(a-c)).
To estimate the magnetostrictive energy, we have used the maximum
theoretical elastic strain value of 0.1\% for both Co and Ni, and
0.24\% for Fe$_{\text{50}}$Co$_{\text{50}}$, obtained from simulation
(see Table\ref{tab:Table}). In addition, the magnetostrictive energy
contribution (E$_{MS}$) was calculated as a function of strain (\%)
in the particles. Fig.\ref{fig:Energy-Phase-plots} show such plots
for Co, Ni, and Fe$_{\text{50}}$Co$_{\text{50}}$, along with the
magnetocrystalline energy, E$_{MC}$, and demagnetization energy,
E$_{DM}$, as a function of number of grains in the nanoparticles.
The calculated values of E$_{MS}$ for maximum strain present in Co,
Ni and Fe$_{\text{50}}$Co$_{\text{50}}$ particles are, $7.7\times10^{3}$,
$1.1\times10^{4}$ and $3.3\times10^{4}$ J/m$^{\text{3}}$, respectively.
It clearly shows that, E$_{MS}$ dominates over other energy terms
(E$_{MC}$ or E$_{DM}$), even when a very small amount of strain
($\sim0.1\%$) is present in these magnetic particles. The dominance
of the magnetostrictive energy, therefore controls the magnetization
direction of the nanoparticles. 

With this reasoning, the difference in the behavior of magnetization
direction for the nanoparticles of Co and Ni versus Fe$_{\text{50}}$Co$_{\text{50}}$
can be understood. Melting under the 9 ns laser pulse and the subsequent
solidification during cooling is associated with large cooling rates,
of the order of $\sim10^{10}$ K/s \cite{Trice07}. One consequence
of this quenching is a residual intrinsic biaxial tensile strain within
the nanoparticles because of the thermal expansion mismatch between
the metal and the SiO$_{\text{2}}$ substrate \cite{krishna08a}.
For an in-plane tensile strain, the magnetization will be perpendicular
to the substrate plane when the $\lambda_{S}$ is negative, as is
the case for Co and Ni. On the other hand, for positive values of
$\lambda_{S}$, the magnetization will be in-plane, as is the case
for Fe$_{\text{50}}$Co$_{\text{50}}$. One point needs to be clarified,
however. A fairly significant number of Fe$_{\text{50}}$Co$_{\text{50}}$
nanoparticles (about 30\%), also show slightly out of plane ($0-45^{o}$)
magnetization. Interestingly, as mentioned above, the TEM analysis
of Fe$_{\text{50}}$Co$_{\text{50}}$ showed large statistical variations
in the number of grains in the same diameter nanoparticles. The magnetocrystalline
anisotropy contribution cannot be completely neglected when the number
of grains is small. For example, the magnetocrystalline anisotropy
energy for 2 grains was calculated to be $\sim1\times10^{4}$ J/m$^{\text{3}}$,
compared to an order of magnitude smaller value of $\leq1\times10^{3}$
J/m$^{\text{3}}$ for 15 grains (anisotropy constant $K_{1}=4.8\times10^{4}$
J/m$^{\text{3}}$ \cite{CullityBook}). The magnetocrystalline energy,
therefore, may compete with the magnetostrictive energy for particles
containing smaller number of grains and orient the magnetization slightly
off from the in-plane direction. The same argument may be applied
to the in-plane magnetization of smaller Ni nanoparticles. Moreover,
some statistical variation in the amount of strain in particles of
the same size, or a variation due to the different number of grains
in the particles, may also be partly responsible for the above observations.

\section*{5. Conclusion}

In conclusion, we have synthesized magnetic nanoparticle arrays of
Co, Ni, and Fe$_{50}$Co$_{50}$ using $ns$ laser-induced self-organization
from ultrathin films deposited on SiO$_{\text{2}}$ surfaces. The
resulting nanoparticles are hemispherical in shape with polycrystalline
microstructure. An extensive study of the orientation of the magnetization
as a function of nanoparticle size was performed using zero-field
MFM. This revealed that the single-domain magnetic nanoparticles of
Co and Ni were primarily oriented out-of-plane. On the other hand,
nanoparticles of Fe$_{50}$Co$_{50}$ were primarily oriented in-plane.
The reason for this difference was attributed to the difference in
the sign of magnetostriction coefficients. Magnetic energy arguments
showed that the magnetostrictive energy dominates among all other
contributions, when some residual tensile strain is present in the
nanoparticles due to the fast cooling process following the ns pulsed
laser irradiation. As a result, metals with negative magnetostriction
coefficient (Co, and Ni) show out-of-plane magnetization while, positive
magnetostriction coefficient materials (Fe$_{50}$Co$_{50}$) showed
in-plane magnetization. This demonstrates a cost-effective, non-epitaxial,
laser-based processing technique for the production of one- and two-dimensional
arrays of magnetic nanoparticles with controlled magnetization directions.

\section*{Acknowledgments}

RK acknowledges support by the National Science Foundation through
CAREER grant NSF-DMI-0449258, and NSF-CMMI-0855949, while RK and AKG
acknowledge grant NSF-DMR-0856707. AKG and HK also acknowledge support
from CMI . HK would also like to acknowledge Vanessa Ramos for her
inputs in the preparation of this manuscript. 

\bibliographystyle{ieeetr}

\begin{flushleft}
\pagebreak{}
\par\end{flushleft}

\begin{table}[H]
\begin{centering}
\begin{tabular}{|c|c|c|c|c|c|c|c|}
\hline 
\multicolumn{2}{|c|}{Property} & SiO$_{2}$ & \multicolumn{2}{c|}{Cobalt} & \multicolumn{2}{c|}{Nickel} & Fe$_{50}$Co$_{50}$\tabularnewline
\hline
\hline 
Thermal expansion coeff. & $\alpha\times10^{-6}$ & 0.55 & \multicolumn{2}{c|}{13} & \multicolumn{2}{c|}{13.4} & 12.4\tabularnewline
\hline 
Melting temperature &  $T_{m}$(K) &  & \multicolumn{2}{c|}{1768} & \multicolumn{2}{c|}{1728} & 1748\tabularnewline
\hline 
Young's modulus & E (GPa) & 72 & \multicolumn{2}{c|}{209} & \multicolumn{2}{c|}{200} & 82.7\tabularnewline
\hline 
Poisson's ratio & $\nu$ & 0.17 & \multicolumn{2}{c|}{0.31} & \multicolumn{2}{c|}{0.31} & 0.31\tabularnewline
\hline 
Yield strength & $\sigma_{y}$ (MPa) & \ensuremath{\infty} & \ensuremath{\infty} & 400 & \ensuremath{\infty} & 250 & \ensuremath{\infty}\tabularnewline
\hline 
Avg. in-plane elastic strain & $\epsilon\left(\%\right)$ &  & 0.10 & 0.04 & 0.11 & 0.5 & 0.24\tabularnewline
\hline
\end{tabular}
\par\end{centering}

\caption{List of material parameters used for finite element simulation and
the average in-plane elastic strain values obtained from the simulation.
The strain reported is the larger of the radial and hoop in-plane
principal strains averaged over the volume of the particle. The yield
strength corresponding to $\infty$ values denote the purely elastic
calculations while the columns with the finite numerical values of
$\sigma_{y}$ correspond to the elastic-perfectly plastic calculations.
Since the $\sigma_{y}$ for the Fe$_{\text{50}}$Co$_{\text{50}}$
alloy was note available, only the elastic result is presented.\label{tab:Table}}

\end{table}

\pagebreak{}

\begin{flushleft}
\begin{figure}[H]
\begin{centering}
\subfloat[]{\includegraphics[width=3in]{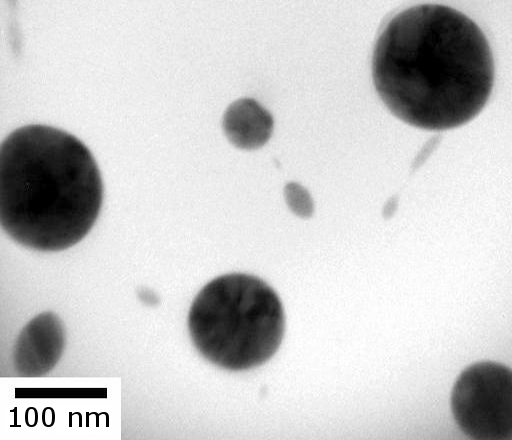} 

}
\par\end{centering}

\begin{centering}
\subfloat[]{\includegraphics[width=3in]{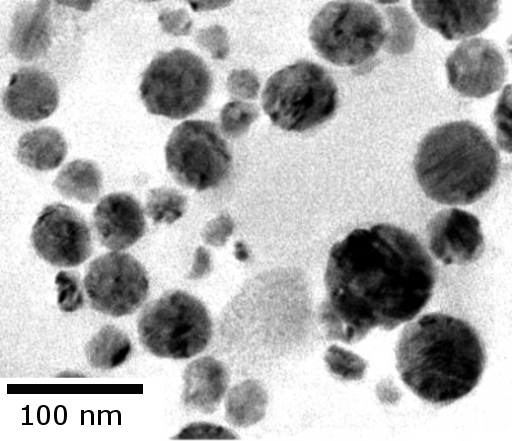}

} 
\par\end{centering}

\begin{centering}
\subfloat[]{\includegraphics[width=3in]{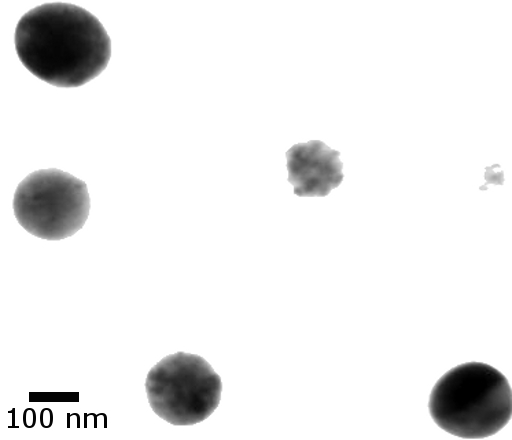}

}
\par\end{centering}

\caption{Representative bright field TEM micrographs for nanoparticle arrays
synthesized by the laser-induced self-organization; (a) Co nanoparticles,
(b) Ni nanoparticles, and (c) Fe$_{\text{50}}$Co$_{\text{50}}$ nanoparticles.
The contrast within each nanoparticle arises from random crystallographic
orientation of multiple grains. Such images were used to generate
statistics on the number of grains as a function of nanoparticle size.
\label{fig:Bright-field-TEM}}

\end{figure}

\par\end{flushleft}

\pagebreak{}

\begin{flushleft}
\begin{figure}[H]
\subfloat[]{\includegraphics[width=3in]{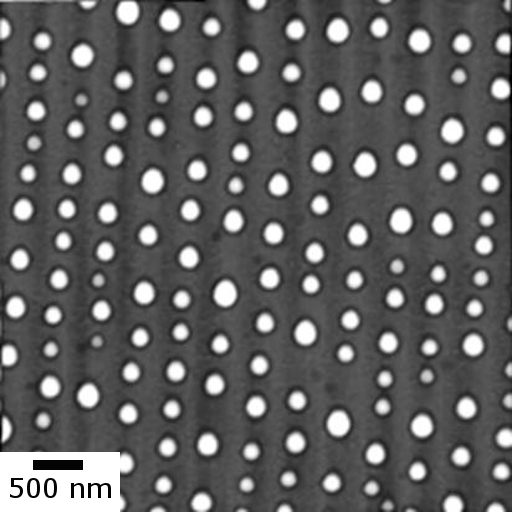} } \subfloat[]{\includegraphics[width=3in]{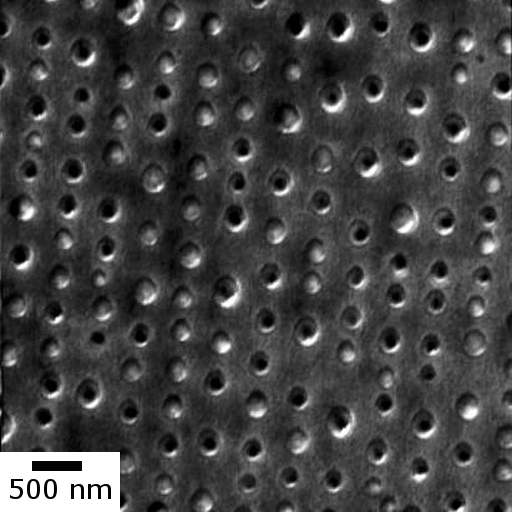}

}

\caption{AFM (a) and zero field MFM (b) images of one dimensional patterned
Co nanoparticles produced by 2-beam pulsed laser interference irradiation
of a 4 nm Co film.\label{fig:Co-AFM-MFM}}

\end{figure}

\par\end{flushleft}

\pagebreak{}

\begin{flushleft}
\begin{figure}[H]
\begin{centering}
\includegraphics[width=4in]{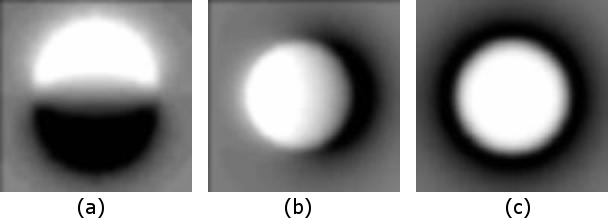}
\par\end{centering}

\caption{The simulated contrast in the MFM images of a single domain ferromagnetic
particle with (a) in-plane ($0^{o}$), (b) at an angle of $45^{o}$
and (c) perpendicular to the plane ($90^{o}$) (taken from ref. \cite{krishna08a}).\label{fig:MFM-simulation}}

\end{figure}

\par\end{flushleft}

\pagebreak{}

\begin{flushleft}
\begin{figure}[H]
\subfloat[]{\includegraphics[width=3in]{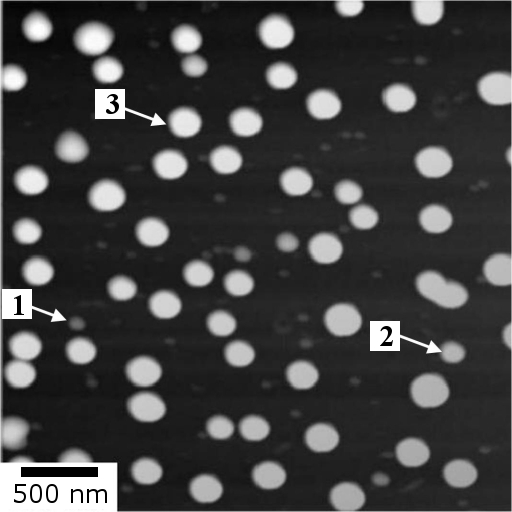}

} \subfloat[]{\includegraphics[width=3in]{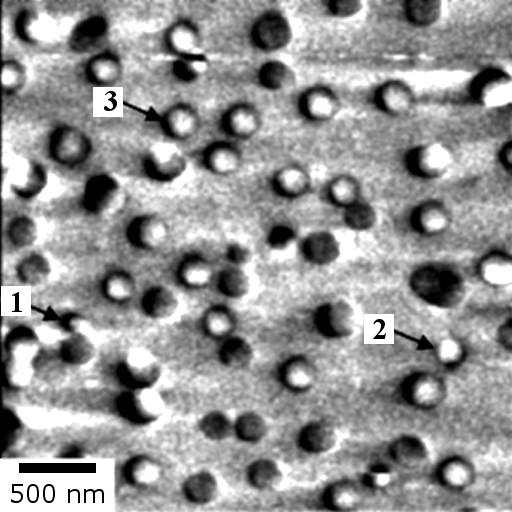}

}

\caption{AFM (a) and zero field MFM (b) images of Ni nanoparticles produced
by single beam pulsed laser irradiation of a 5 nm Ni film. The nanoparticles
marked as \#s 1, 2, and 3 in the AFM image (Fig. (a)) are $75$ nm,
$135$ nm and $200$ nm diameter, respectively. The corresponding
MFM image {[}Fig. (b){]} indicates the magnetization directions with
respect to the substrate plane at $0^{o}$, $45^{o}$ and $90^{o}$,
respectively.\label{fig:Ni-AFM-MFM}}

\end{figure}

\par\end{flushleft}

\pagebreak{}

\begin{flushleft}
\begin{figure}[H]
\subfloat[]{\includegraphics[width=3in]{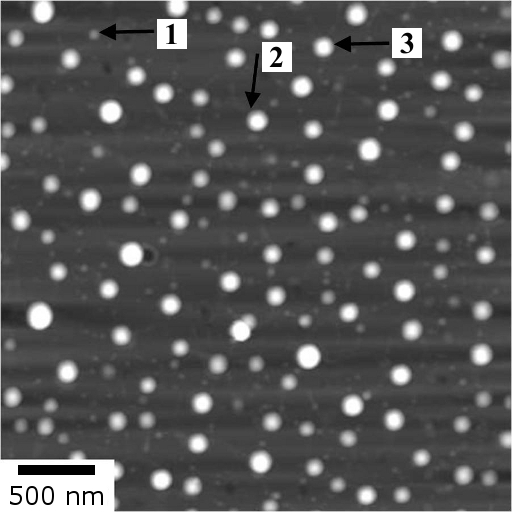}

} \subfloat[]{\includegraphics[width=3in]{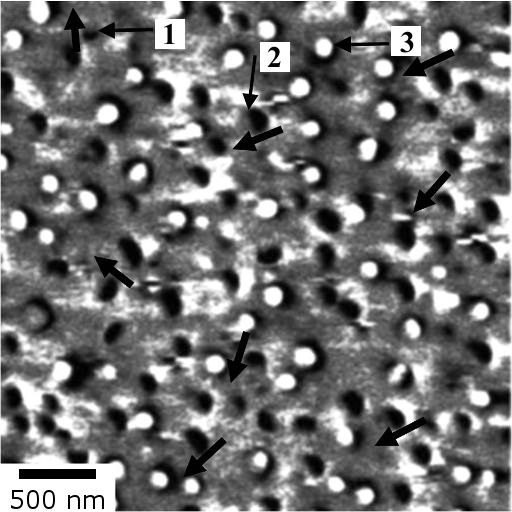}

}

\caption{AFM (a) and the zero-field MFM (b) images of Fe$_{\text{50}}$Co$_{\text{50}}$
nanomagnets produced from a 4 nm film by pulsed laser irradiation.
The nanoparticles indicated as \#s 1 and 2 in the AFM image (Fig.
(a)) are nm and $150$ nm diameters; the corresponding MFM image (Fig.
(b)) indicates that both have magnetization direction in the substrate
plane ($0^{o}$), while another 150 nm diameter particle ( \# 3) is
aligned at $\sim45^{o}$ to the substrate. The bold arrows in Fig.
(b) indicate the in-plane random orientations of the other nanoparticles.
\label{fig:FeCo-AFM-MFM}}

\end{figure}

\par\end{flushleft}

\pagebreak{}

\begin{flushleft}
\begin{figure}[H]
\begin{centering}
\subfloat[]{\includegraphics[height=2.5in]{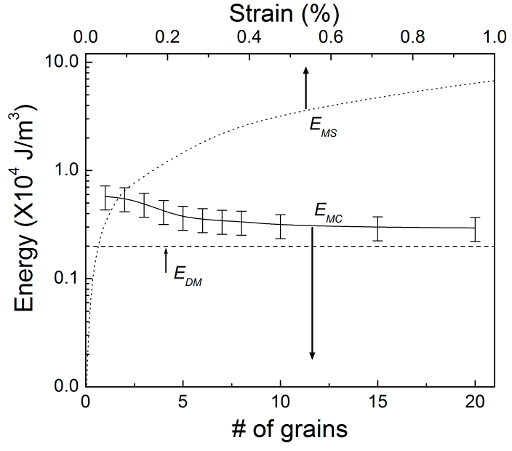}

}\vspace{-0.1in}
\par\end{centering}

\begin{centering}
\subfloat[]{\includegraphics[height=2.5in]{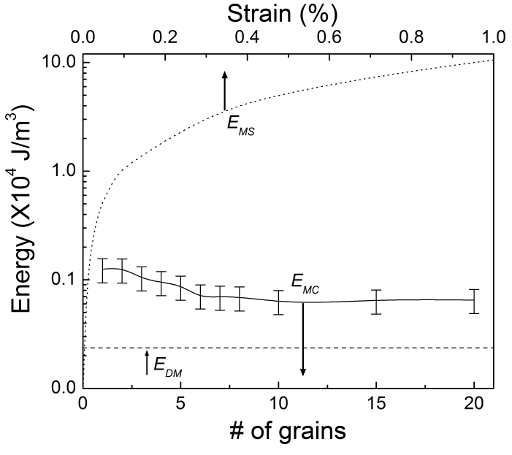}

}\vspace{-0.1in}
\par\end{centering}

\begin{centering}
\subfloat[]{\includegraphics[height=2.5in]{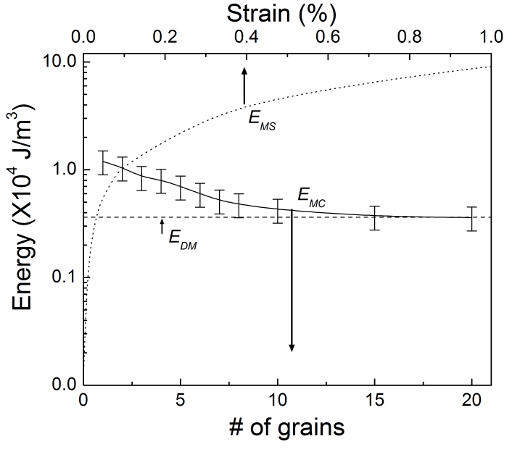}

}\vspace{-0.1in}
\par\end{centering}

\caption{The comparison of different magnetic energies for the three different
materials. The magnetocrystalline energy (E$_{MC}$) is shown as a
function of number of grains\emph{,} and the magnetostrictive energy
(E$_{MS}$) as a function of strain (\%). The demagnetization energy
(E$_{DM}$) is also shown. Fig (a) corresponds to Co, Fig. (b) to
Ni and, Fig. (c) to Fe$_{\text{50}}$Co$_{\text{50}}$. In each of
the three cases a small amount of residual strain (\textasciitilde{}0.1\%)
is sufficient to make the E$_{MS}$ dominate over E$_{MC}$ and E$_{DM}$.
\label{fig:Energy-Phase-plots}}

\end{figure}

\par\end{flushleft}
\end{document}